\documentclass[twoside,a4paper,11pt]{proceedings}
\usepackage{graphicx}
\usepackage{hyperref}
\usepackage{movie15}
\usepackage{natbib}
\newcommand{\apj}{ApJ}
\newcommand{\apjl}{ApJL}
\newcommand{\aap}{A\&A}
\newcommand{\mnras}{MNRAS}
\newcommand{\kms}{\rm ~km~s^{-1}}
\begin{document}
\pagenumbering{arabic}
\pagestyle{myheadings}
\thispagestyle{empty}
\vspace*{-1cm}
{\flushleft\includegraphics[width=3cm,viewport=0 -30 200 -20]{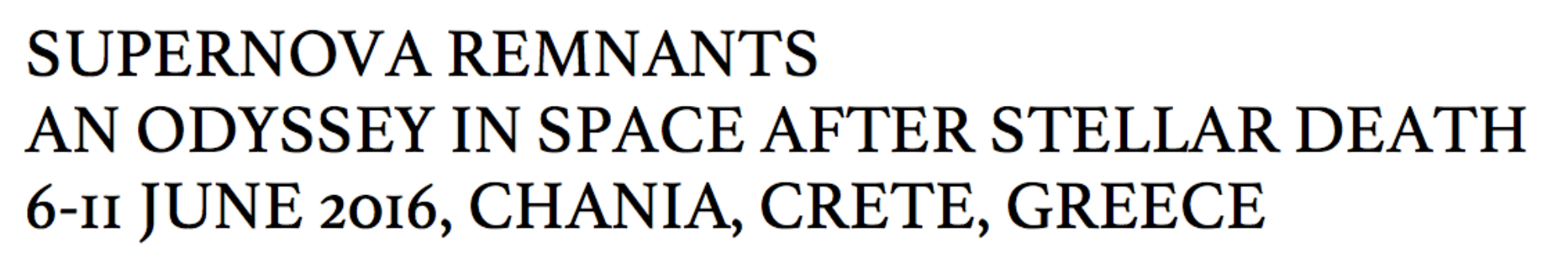}}
\vspace*{0.2cm}
\begin{flushleft}
{\bf {\LARGE
Time dependent diffusive shock acceleration and its application to middle aged supernova remnants
}\\
\vspace*{1cm}
Xiaping Tang$^{1}$ and 
Roger A. Chevalier$^{2}$
%
}\\
\vspace*{0.5cm}
%
$^{1}$
Max Planck Institute for Astrophysics, Karl-Schwarzschild-Str. 1,
D-85741 Garching, Germany\\
$^{2}$
Department of Astronomy, University of Virginia, P.O. Box 400325, Charlottesville, VA 22904-4325, USA

%
\end{flushleft}
\markboth{
Time dependent DSA 
}{
Xiaping Tang \& Roger A. Chevalier
}
\thispagestyle{empty}
\vspace*{0.4cm}
\begin{minipage}[l]{0.09\textwidth}
\ 
\end{minipage}
\begin{minipage}[r]{0.9\textwidth}
\vspace{1cm}
\section*{Abstract}{\small
Recent $\gamma$-ray observations show that middle aged supernova remnants (SNRs) interacting with molecular clouds (MCs) can be sources of both GeV and TeV emission. Based on the MC association, two scenarios have been proposed to explain the observed $\gamma$-ray
emission. In one, energetic cosmic ray (CR) particles escape from the SNR
and then illuminate nearby MCs, producing $\gamma$-ray emission, while the other involves direct
interaction between the SNR and MC. In the direct interaction scenario, re-acceleration of pre-existing CRs in the ambient medium is investigated while particles injected from the thermal pool are neglected in view of the slow shock speeds in middle aged SNRs. However, standard diffusive shock acceleration (DSA) theory produces a steady state particle spectrum that is too flat compared to observations, which suggests that the high energy part of the observed spectrum has not yet reached a steady state. We derive a time dependent DSA solution in the test particle limit for re-acceleration of pre-existing CRs case and show that it is capable of reproducing the observed $\gamma$-ray emission in SNRs like  IC 443 and W44, in the context of a MC interaction model. We also provide a simple physical picture to understand the time dependent DSA spectrum. A spatially averaged diffusion coefficient around the SNR can be estimated through fitting the $\gamma$-ray spectrum. The spatially averaged diffusion coefficient in middle aged SNRs like IC 443 and W44 is estimated to be $\sim 10^{25}\rm cm^2s^{-1}$ at $\sim \rm 1GeV$, which is between the Bohm limit and interstellar value.

\vspace{10mm}
\normalsize}
\end{minipage}

\section{Introduction}
Recent observations from both space-based GeV observatories and ground-based TeV observatories reveal $\gamma$-ray emission from several middle aged supernova remnants (SNRs) which are interacting with molecular clouds (MCs), e.g. W44 \citep{Abdo10a,Giuliani11,Uchiyama12}, IC443 \citep{Albert07,Acciari09,Abdo10b}, W28 \citep{Aharonian08,Abdo10c,Hanabata14} and W51C \citep{Abdo09,Aleksic12}. The observed $\gamma$-ray emission from these objects shows a smooth transition from the GeV band to the TeV band (if detected) and they all peak in the GeV band. The characteristic $\pi^0$-decay signature identified in IC 443 and W44 \citep{Giuliani11,Ackermann13} provides possible direct evidence for cosmic ray (CR) proton acceleration in SNRs, making middle aged SNRs detected in $\gamma$-rays an important class of objects for understanding  CR acceleration in SNRs.

So far two scenarios have been proposed to explain the observed spatial correlation between the $\gamma$-ray emission region and the MC region with hadronic origin: one is the escaping scenario \citep{Gabici09,Fujita09,LC10,Ohira11}, which focuses on the CR particles that escaped from the remnant, and the other one is the direct interaction scenario \citep{bykov00,Uchiyama10,Inoue10,TC14,TC15,Cardillo16}, which studies the CR particles accelerated at the remnant. In the escaping scenario, CR particles having escaped from the remnant together with the pre-existing CR in the ambient medium illuminate the nearby MCs, producing the $\gamma$-ray emission. A collision between the SNR and MC is not necessarily required and the resulting  $\gamma$-ray emission can be external to the remnant. A good example for this scenario might be W28. In the direct interaction scenario, a collision between the remnant and MC occurs, as indicated by both observation and theory. Shocked clouds are identified in middle aged SNRs and are spatially coincident with the $\gamma$-ray emission region  \citep[e.g.,][]{Abdo10a,Abdo10b,uchiyama11,Nicholas12}. Besides, the MC interaction creates a region with high density which then becomes an ideal site for $\pi^0$-decay emission.  

Diffusive shock acceleration (DSA) is believed to be the particle acceleration mechanism in most astrophysical environments involving shock waves \citep[e.g.,][]{Bell78,B&E87}. The theory naturally produces a power law spectrum of energetic particles in a steady state which is close to the observed CR spectrum after taking into account  propagation effects. DSA is considered to be the most plausible particle acceleration mechanism at the SNR shock front for both scenarios.

Here, we limit our discussion to the direct interaction scenario. In this scenario, re-acceleration of pre-existing CRs is taken into account, while particle injection through the thermal pool is neglected in view of the slow radiative shock \citep{B&C82,Uchiyama10,TC14}. If this picture is correct, then it implies a transition of seed particles for CR acceleration in SNRs from  injected seed particles to pre-existing CRs.  When does this transition occur and how would it affect the accelerated CR spectrum in SNRs could be interesting problems for future study.
 
The standard DSA theory produces too flat a steady state particle spectrum compared to that indicated by observations, so it has been suggested that the observed particle spectrum in the energy range of interest has not reached a steady state yet. An exponential cutoff has been explored in \cite{Uchiyama10} and \cite{TC14} as the high energy cutoff for  MC interaction models. It is found that for remnants like IC 443, W28 and W51C with TeV detection, the exponential cutoff tends to underproduce the TeV emission. In \cite{TC15}, we developed a time dependent DSA solution in the test particle limit for the case of re-acceleration of pre-existing CRs and show that the time dependent solution in combination with the MC interaction model in \cite{TC14} is capable of explaining the observed $\gamma $-ray emission in remnants like W44 and IC 443 from GeV to TeV bands.

In the following section, we first discuss  the properties of the time dependent DSA solution  and then provide a simple physical picture to understand the spectral shape of the time dependent DSA solution.



\section{Time dependent DSA solution}
Assume that the pre-existing CRs around the SNR have the same spectrum as that observed locally,
which shows a broken power law. Based on calculations in \cite{TC15} for the time dependent DSA spectrum, there is a critical momentum below which the spectral shape is consistent with the steady state spectrum, and above which the spectral shape simply follows the input CR spectrum with possible hardening due to energy dependent diffusion (see figure 1-4 in \cite{TC15} for details).

In the Appendix of \cite{TC15}, we provided a simple phenomenological way to understand the  time dependent DSA spectrum for energy independent diffusion. Here, we discuss the physical picture in more detail and then discuss  possible implications of such a picture. In DSA, energetic particles bounce back and forth around the shock front due to self generated magnetic turbulence \citep[e.g.][]{Bell78}. Every time a particle moves across the shock discontinuity it gains some energy and every time the particle enters the downstream region it has chance to escape from the DSA site. According to \cite{Drury83}, the momentum gain rate for a particle undergoing DSA is 
\begin{equation}
\frac{dp}{dt}=\frac{U_1-U_2}{3}\, \left( \frac{\kappa_1}{U_1} +\frac{\kappa_2}{U_2}\right)^{-1}  p,
\label{energy_gain}
\end{equation} 
where $U$ is the flow velocity, $\kappa$ is the diffusion coefficient and subscripts 1 and 2 are for physical quantities in the upstream and downstream regions, respectively. 
The probability for a particle to leave the acceleration site at downstream region is given by $P=4U_2/v$ \citep{Drury83}. 

For a particle with initial momentum $p_i$, after $n$ cycles of acceleration the particle momentum becomes (See e.g. \cite{Drury83} for detail)
\begin{equation}
p_n\sim \prod^n_{k=1}\left[1+\frac{4(U_1-U_2)}{3v_k}\right]p_i,
\end{equation}
leading to
\begin{equation}
 {\rm ln}(p_n/p_i)\sim\frac{4(U_1-U_2)}{3}\sum^n_{k=1}\frac{1}{v_k}.
\end{equation}  
The probability for a particle to stay at the acceleration site after $n$ cycles of back and forth motion is 
\begin{equation}
P_n\sim\prod^n_{k=1} \left( 1-\frac{4U_2}{v_k}\right),
\end{equation}
leading to
\begin{equation}
\ln P_n  \sim  -4U_2\sum^n_{k=1}\frac{1}{v_k}
	\approx -\frac{3U_2}{U_1-U_2} {\rm ln}(p_n/p_i).
	\label{escape_probability}
\end{equation}
For a strong shock $P_n\approx p_i/p_n$.

Next consider the acceleration of a group of particles with momentum spectrum $N(p)$. Based on the conservation of particle number, the new particle spectrum $\overline{N}(p)$ is related to the input particle spectrum $N(p)$ by
 \begin{equation}
\overline{N}(p_n)dp_n=N(p_i)P_ndp_i,
\label{conservation}
\end{equation}
where $p_i$ and $p_n$ are the particle momentum before and after the DSA respectively. $P_n$ is the probability for a particle to stay at the DSA site after $n$ cycles of acceleration and equals  $p_i/p_n$ according to eq.\  \ref{escape_probability}. For constant $U$ and $\kappa$, the energy gain rate $dp/dt$ given by eq.\  \ref{energy_gain} is proportional to the particle momentum $p$, which implies that $dp_n/dp_i=p_n/p_i$. Now eq.\  \ref{conservation} becomes $\overline{N}(p_n)p_n^2\approx N(p_i)p_i^2$, which means in the particular $\log[N(p)p^2]-\log (p)$ plane the whole DSA process simply works like a horizontal shift in the $\log(p)$ axis. The amount of shift is determined by
\begin{equation}
{\rm ln}\left( \frac{p_n}{p_i}\right)=\frac{U_1-U_2}{3}\, \left( \frac{\kappa_1}{U_1} +\frac{\kappa_2}{U_2}\right)^{-1}  t,
\label{shift}
\end{equation}
which is linearly proportional to $t$ in the $\log(p)$ axis.

\begin{figure}[htb]
\center
\includegraphics[width=0.8\textwidth]{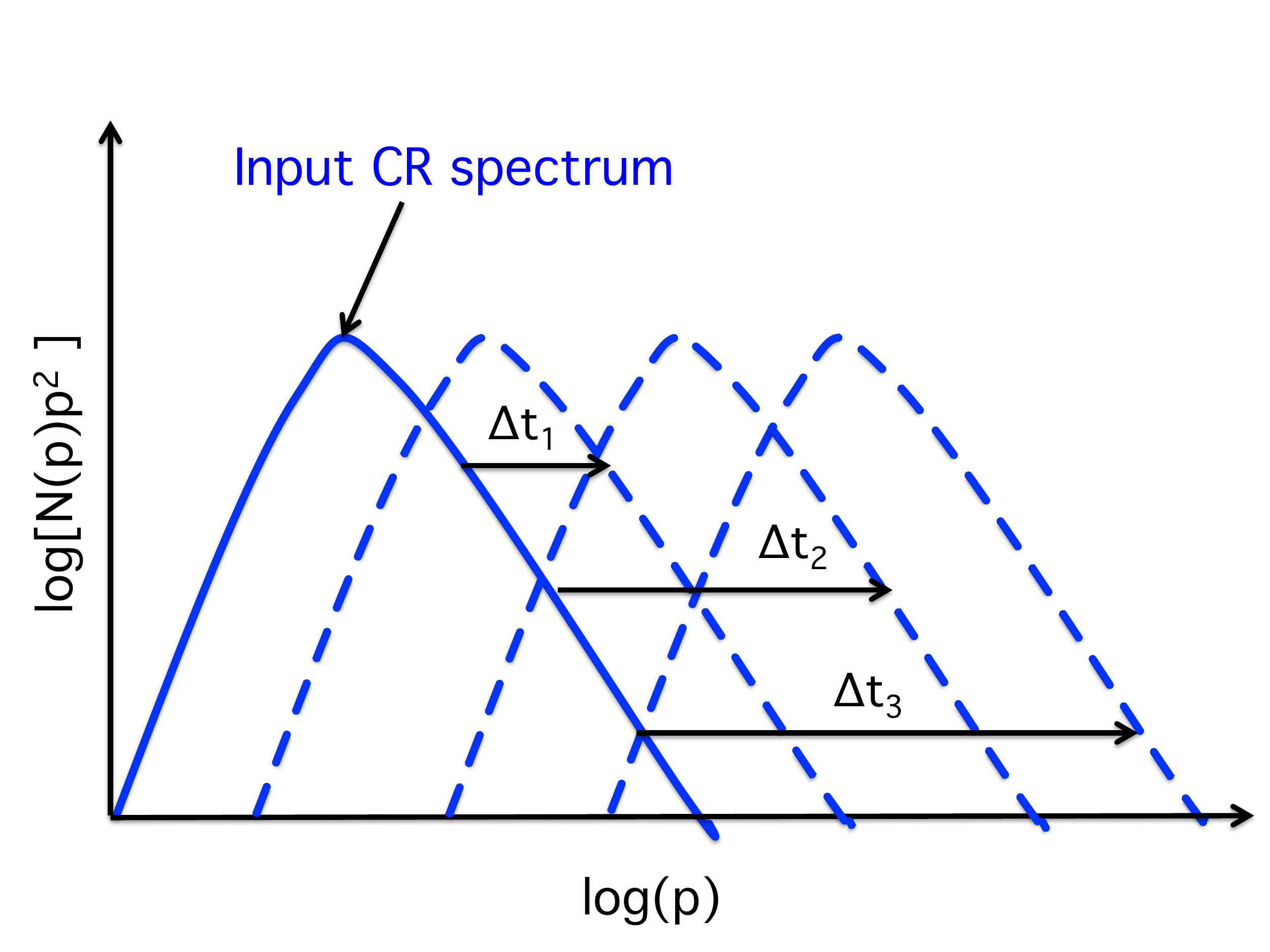} 
\caption{Schematic figure for the time dependent DSA solution.}
\label{schematic}
\end{figure}

In the case of re-acceleration of pre-existing CRs in a SNR, energetic particles with the same input CR spectrum are being injected into the system all the time. Particles injected at different times will be shifted by different amounts in the $\log(p)$ axis due to different acceleration times, as shown in Fig \ref{schematic}. As a result, the accumulated particle spectrum at the shock front is determined by the sum of a group of input CR spectra with different amounts of shift. Since the amount of shift in $\log(p)$ axis is linearly proportional to $t$ according to eq.\  \ref{shift}, all the shifted spectra have the same weight for the sum in $\log[N(p)p^2]-\log (p)$ plane. 
Now we can understand the time dependent DSA spectrum presented in \cite{TC15}. Because our input CR spectrum follows roughly a broken power law, the accumulated CR spectrum at the shock front  contains three parts according to the above discussion. In the low energy and high energy parts the accumulated particle spectrum maintains the shallow and steep power law in the input CR spectrum, respectively, as the shifted spectra all share the same power law index. At intermediate energies the accumulated particle spectrum shows a plateau which is mainly due to the break in the input CR spectrum. 

\section{Discussion}
If the observed $\gamma$-ray emission from those middle aged SNRs interacting with MCs is the result of time dependent DSA,  according to eq.\  \ref{shift} the extension of the plateau region in the $\gamma$-ray spectrum reflects the logarithmic ratio $\ln(p_f/p_i)$ between the initial and final momentum. Thus it can be used to estimate the spatially averaged diffusion coefficient around the SNRs. In SNRs, the spatially averaged diffusion coefficient in the upstream region $\kappa_1$ is much larger than that in the downstream region $\kappa_2$, as $\kappa_1$ gradually approaches the interstellar value when particles move far away from the shock front. Hence, according to eq.\ \ref{shift} we have  
\begin{equation}
\kappa_1 \approx\frac{(U_1-U_2)U_1t}{3\ln (p_f/p_i)}\approx 10^{25}{\rm cm^2s^{-1}}\frac{r-1}{r\ln(p_f/p_i)}\left(\frac{U_1}{100\kms}\right)^2\left(\frac{t}{10^4~ \rm yrs}\right),
\end{equation}
where $r$ is the shock compression ratio and $r=4$ for strong shocks. In middle aged SNRs $\ln(p_f/p_i)$ is of order  unity based on the observed $\gamma$-ray spectrum. Hence the spatially averaged diffusion coefficient in these middle aged SNRs at the energy range around the plateau region ($\sim \rm GeV$) is approximately $10^{25}\rm cm^2s^{-1}$. The number is close to the value derived in \cite{TC15} through detailed fitting and is found to be between the Bohm limit and the interstellar value.  

\small  
%
\section*{Acknowledgments}   
%
We are very grateful to P. Boumis and the meeting organizers for arranging a very stimulating conference.


\end{document}